\documentclass[preprint,11pt]{elsarticle}
\usepackage{amssymb}
\usepackage{mathrsfs}
\usepackage{epic,eepic,epsf,epsfig}
\usepackage{graphicx}
\usepackage{amsfonts,srcltx,mathrsfs}
\usepackage{amsmath}
\usepackage{amsbsy}
\usepackage{amsfonts}
\usepackage{color}
\usepackage{array}

\usepackage{amssymb}
\usepackage{amsthm}
\newtheorem{theorem}{Theorem}
\newtheorem{lem}{Lemma}
\newtheorem{cor}{Corollary}
\newtheorem{clm}{Claim}
\newdefinition{rmk}{Remark}
\newdefinition{defi}{Definition}
\newproof{pf}{Proof}
\newdefinition{prop}{Proposition}

\def\f{\noindent}

\def\demo{{\bf Proof.}\hskip10pt}

\journal{}

\begin{document}

\begin{frontmatter}

\title{Fault diagnosability of data center networks}

\author[bjtu]{Mei-Mei Gu}
\ead{12121620@bjtu.edu.cn,}

\author[bjtu]{Rong-Xia Hao\corref{cor1}}
\ead{rxhao@bjtu.edu.cn,}

\author[fjnu]{Shuming Zhou}
\ead{zhoushuming@fjnu.edu.cn,}

\address[bjtu]{Department of Mathematics, Beijing Jiaotong University, Beijing, 100044, China}
\address[fjnu]{School of Mathematics and Computer Science, Fujian Normal University, Fuzhou, Fujian 350108, China}

\cortext[cor1]{Corresponding author}

\begin{abstract}
The data center networks $D_{n,k}$, proposed in 2008, has many desirable features such as high
network capacity.
A kind of generalization of diagnosability for network $G$ is $g$-good-neighbor diagnosability which is denoted by $t_g(G)$.
Let $\kappa^g(G)$ be the $R^g$-connectivity. Lin et. al. in [IEEE Trans. on Reliability, 65 (3) (2016) 1248--1262] and Xu
et. al in [Theor. Comput. Sci. 659 (2017) 53--63] gave the same problem independently that: the relationship between the $R^g$-connectivity $\kappa^g(G)$ and $t_g(G)$ of
a general graph $G$ need to be studied in the future. In this paper, this open problem is
solved for general regular graphs.
We firstly establish the relationship of $\kappa^g(G)$ and $t_g(G)$, and obtain that $t_g(G)=\kappa^g(G)+g$ under some conditions. Secondly, we obtain the $g$-good-neighbor diagnosability
of $D_{k,n}$ which are $t_g(D_{k,n})=(g+1)(k-1)+n+g$ for $1\leq g\leq n-1$ under the PMC model
and the MM model, respectively. Further more, we show that $D_{k,n}$ is tightly super $(n+k-1)$-connected for $n\geq 2$ and $k\geq 2$ and we also prove that the largest connected component of the survival graph contains almost all of the remaining vertices in $D_{k,n}$ when $2k+n-2$ vertices removed.
\end{abstract}


\end{frontmatter}

\section{Introduction}

The study of interconnection networks has been an important
research area for parallel and distributed computer systems.
A network can be modeled as a graph, in which vertices and
edges correspond to processors and communication links, respectively.
Network reliability is one of the major factors in designing the topology of an interconnection network.
With the rapid development of multiprocessor systems,
processor failure is inevitable along with the number of processors increasing.
The process of identifying all the faulty
units in a system is called as {\it system-level diagnosis}.
For the purpose of self-diagnosis of a system, a number of models have been proposed for diagnosing faulty processors in a network. Among the proposed models,
PMC model~\cite{P} and comparison model (MM model)~\cite{mm81} are widely used.
In the PMC model, every processor can test the processor that is adjacent to it and
only the fault-free processor can guarantee reliable outcome.
In the MM model, to diagnose the system, a processor sends the same task to one pair of its neighbors, and then compares their responses.
A system is said to be {\it $t$-diagnosable} if all faulty units can be identified provided the number of faulty units present does not exceed $t$.
The {\it diagnosability} is the maximum number of
faulty processors which can be correctly identified. In 2005, Lai et al.~\cite{l05} introduced a restricted diagnosability of the system called {\it conditional diagnosability} by assuming that it is impossible that all neighbors of one vertex are faulty simultaneously. The diagnosabilities and conditional diagnosabilities of many networks are studied in literatures ~\cite{C1}-\cite{C3},~\cite{F02}-\cite{H1},~\cite{H2},~\cite{Hc13}-\cite{Hk13},
\cite{lz12},~\cite{lz14},~\cite{WD},~\cite{Z12} etc.
Inspired by this concept, Peng et al.~\cite{P12} then proposed the {\it $g$-good-neighbor
diagnosability}, which requires every fault-free vertex has at least $g$ fault-free neighbors.

\begin{defi}\label{defi1}
A fault set $F\subseteq V(G)$ is a {\it $g$-good-neighbor faulty set}
if $|N_{G}(v)\cap (V(G)\setminus F)|\geq g$ for every vertex $v\in V(G)\setminus F$.
A {\it $g$-good-neighbor cut} of a graph $G$ is a $g$-good-neighbor
faulty set $F$ such that $G-F$ is disconnected.
For an arbitrary graph $G$, $g$-good-neighbor cuts do not always exist for some $g$.
A graph $G$ is called an {\it $R^g$-graph} if it contains at least one $g$-good-neighbor cut.
For an $R^g$-graph $G$, the minimum cardinality of $g$-
good-neighbor cuts is said to be the {\it $R^g$-connectivity} of $G$,
denoted by $\kappa^g(G)$. The parameter $\kappa^1(G)$ is equal to {\it extra connectivity}
$\kappa_1(G)$ which is proposed by F\'abrega and Fiol \cite{F96},
where $\kappa_k(G)$ is the cardinality of a minimum set $S\subseteq V(G)$ such
that $G-S$ is disconnected and each component of $G-S$ has at least $k+1$ vertices.
\end{defi}

\begin{defi}\label{defi2}
A system $G=(V,E)$ is {\it $g$-good-neighbor $t$-diagnosable}
if $F_1$ and $F_2$ are distinguishable (the definition of distinguishable is in Section 2),
for each distinct pair of $g$-good-neighbor
faulty sets $F_1$ and $F_2$ of $V$ with $|F_1|\leq t$ and $|F_2|\leq t$.
The {\it $g$-good-neighbor diagnosability} $t_g(G)$ of a graph
$G$ is the maximum value of $t$ such that $G$ is $g$-good-neighbor $t$-diagnosable.
\end{defi}

The classical diagnosability relies on an assumption that all neighbors of each vertex in a parallel system can potentially fail at the same time. But the $g$-good-neighbor diagnosability is superior to the classical diagnosability in terms of measuring diagnosability for large-scale parallel systems.
The problem of determining the $g$-good-neighbor diagnosability for $g=1,2$
of numerous networks, for examples, see~\cite{W15} and~\cite{W16}, has received much attention in recent years. But little is known about $t_g(G)$ with a general non-negative integer $g$ for networks except for  hypercubes, $k$-ary $n$-cubes etc. Peng et al.~\cite{P12} showed that the $g$-good-neighbor diagnosability of the $n$-dimensional hypercube $Q_n$ under the PMC model is $2^g(n-g)+2^g-1$ for $0\leq g\leq n-3$.
Yuan et al.~\cite{Y15} and~\cite{Y16} studied the $g$-good-neighbor diagnosability of the $k$-ary $n$-cubes ($k\geq4$) and $3$-ary $n$-cubes,
respectively, under the PMC model and MM model.
Wang and Han~\cite{WH16} determined the $g$-good-neighbor diagnosability of the $n$-dimensional hypercube $Q_n$ under the MM model.

Xu et al.~\cite{lzh} and Lin et al.~\cite{xz16} gave the same problem independently that {\bf the relationship between the $R^g$-connectivity $\kappa^g(G)$ and $t_g(G)$ of a general graph $G$ need to be studied in the future}.

In this paper, we firstly study the relation between $g$-good-neighbor diagnosability and
$R^g$-connectivity for regular graphs and obtain the following Theorem~\ref{th1}.
Secondly, we prove that $D_{k,n}$ is tightly super $(n+k-1)$-connected for $n\geq 2$ and $k\geq 2$ and we also prove that the largest connected component of the survival graph contains almost all of the remaining vertices in $D_{k,n}$ when almost $2k+n-2$ vertices are removed. Thirdly, we obtain that the $g$-good-neighbor diagnosability
of $D_{k,n}$ which are $t_g(D_{k,n})=(g+1)(k-1)+n+g$ for $1\leq g\leq n-1$ under the PMC model and the MM model, respectively. As direct corollaries, the $g$-good-neighbor diagnosability
of the $(n,k)$-star networks $S_{n,k}$ and the $(n,k)$-arrangement graphs $A_{n,k}$ are obtained.

\begin{theorem}\label{th1}
Let $n$, $g$ and $N$ be non-negative integers.
Let $G$ be an $n$-regular connected $R^g$-graph with order $N$.
Suppose $G$ has a complete subgraph $K_m$ of order $m$, where $m\leq n-1$.
Let $\kappa^g(G)$ be the $R^g$-connectivity of $G$. If $G$ satisfies the conditions (1) and (2) under the PMC model;
or $G$ satisfies the conditions (1),(2) and (3) under the MM model.
\begin{enumerate}
\item [{\rm (1)}] there exists a minimum $g$-good-neighbor cut $T$ such that $G-T$ has exactly two components, one of which is isomorphic to $K_{g+1}$, where $g\leq m-1$;
\item [{\rm (2)}] $N\geq 2\kappa^g(G)+3\kappa^1(G)+2g-n-1$;
\item [{\rm (3)}] for any $F\subseteq V(G)$ and $|F|\leq \kappa^1(G)$, $G-F$ is either connected;
or has two components, one of which is a trivial component;
or has two components, one of which is an edge;
or has three components, two of which are trivial components.
\end{enumerate}

\f Then, $t_g(G)=\kappa^g(G)+g$ for $1\leq g\leq m-1$.

\end{theorem}

The remainder of this paper is organized
as follows. Section 2 introduces some necessary notations and basic lemmas.
Our main results are given in Section 3. As applications of our main result,
Section 4 concentrates on the $g$-good-neighbor diagnosability of three kinds of graphs:
data center networks  $D_{n,k}$, the $(n,k)$-star networks $S_{n,k}$,
the $(n,k)$-arrangement graphs $A_{n,k}$.
Section 5 concludes the paper.

\section{Preliminaries}
In this section, we give some terminologies and notations
of combinatorial network theory.
We follow~\cite{x03} for terminologies and notations not defined here.

We use a graph, denoted by $G=(V(G),E(G))$, to represent an interconnection network,
where a vertex $u\in V(G)$ represents a processor and an edge $(u,v)\in E(G)$
represents a link between vertices $u$ and $v$.
Two vertices $u$ and $v$ are {\it adjacent} if $(u,v)\in E(G)$, the vertex $u$ is
called a neighbor of $v$, and vice versa.
For a vertex $u\in V(G)$, let $N_{G}(u)$ denote
a set of vertices in $G$ adjacent to $u$. The cardinality $|N_{G}(u)|$ represents the {\it degree} of $u$ in $G$, denoted by $d_{G}(u)$ (or simply $d(u)$), $\delta(G)$ the {\it minimum degree} of $G$.
For a vertex set $U \subseteq V(G)$, the {\it neighborhood}
of $U$ in $G$ is defined as $N_G(U)=\bigcup\limits_{v\in U}N_{G}(v)-U$.
If $|N_{G}(u)|=k$ for any vertex in
$G$, then $G$ is {\it $k$-regular}. Let $G$ be a connected graph, if $G-S$ is still connected for any $S\subseteq V(G)$ with $|S|\leq k-1$, then $G$ is {\it $k$-connected.}
A subset $S\subseteq V(G)$ is a {\it vertex cut} if $G-S$ is disconnected.
The {\it connectivity} of a graph $G$, denoted by $\kappa(G)$,
defined as the minimum number of vertices whose removal results in
a disconnected or trivial graph.
A $k$-regular graph is {\it loosely super $k$-connected} if any one of its minimum vertex cuts is a
set of the neighbors of some vertex. If, in addition, the deletion of a minimum vertex cut results in a graph with two components (one of which has only one vertex), then the graph is {\it tightly super $k$-connected}.
A graph $H$ is a {\it subgraph} of a graph $G$ if $V(H)\subseteq V(G)$
and $E(H)\subseteq E(G)$.
The {\it components} of a graph $G$ are its
maximally connected subgraphs. A component is {\it trivial} if it has only
one vertex; otherwise, it is {\it nontrivial}.

To diagnose faults, a number of tests are performed on vertices.
The collection of all test results is called a {\it syndrome}.
Let $F$ be a subset of $V(G)$. $F$ is said to be {\it compatible with}
a syndrome $\sigma$ if $\sigma$ can
arise from the circumstance that all vertices in $F$ are faulty and all
vertices in $V(G)\setminus F$ are fault free. A system is said to be {\it diagnosable} if,
for every syndrome $\sigma$, there is a unique $F\subseteq V(G)$ such that $F$ is
compatible with $\sigma$.
Let $\sigma_{F}=\{\sigma: \sigma$ is compatible with $F\}$.
Two distinct subsets $F_1,F_2\subseteq V(G)$ are said to be {\it indistinguishable}
if and only if $\sigma_{F_1}\cap\sigma_{F_2}\neq\emptyset$;
otherwise, $F_1,F_2$ are said to be {\it distinguishable}.
The {\it symmetric difference} of $F_1\subseteq V(G)$ and
$F_2\subseteq V(G)$ is defined as the set $F_1\Delta F_2=(F_1\setminus F_2)\cup(F_2\setminus F_1)$.

The following two lemmas characterize a graph for $g$-good-neighbor $t$-diagnosable
under the PMC model and the MM model, respectively.

\begin{lem}{\rm(\cite{sd92,Y15})}\label{lem-pmc}
A system $G=(V,E)$ is $g$-good-neighbor $t$-diagnosable
under the PMC model if and only if there is an edge $(u,v)\in E$ with
$u\in V\setminus(F_1\cup F_2)$ and $v\in F_1\Delta F_2$ for each distinct
pair of $g$-good-neighbor faulty sets $F_1$ and $F_2$ of $V$ with $|F_1|\leq t$ and $|F_2|\leq t$.
\end{lem}

\begin{lem}{\rm(\cite{D84,Y15})}\label{lem-mm}
A system $G=(V,E)$ is $g$-good-neighbor $t$-diagnosable
under the MM model if and only if for each distinct pair of $g$-good-neighbor
faulty sets $F_1$ and $F_2$ of $V$ with $|F_1|\leq t$ and $|F_2|\leq t$ satisfies one of the following conditions.
\begin{enumerate}
\item [{\rm (1)}] There are two vertices $u,w\in V\setminus (F_1\cup F_2)$
and there is a vertex $v\in F_1\Delta F_2$ such that $(u,v)\in E$ and $(u,w)\in E$.
\item [{\rm (2)}] There are two vertices $u,v\in F_1\setminus F_2$
and there is a vertex $w\in V\setminus (F_1\cup F_2)$ such that $(u,w)\in E$ and $(v,w)\in E$.
\item [{\rm (3)}] There are two vertices $u,v\in F_2\setminus F_1$
and there is a vertex $w\in V\setminus (F_1\cup F_2)$ such that $(u,w)\in E$ and $(v,w)\in E$.
\end{enumerate}

\end{lem}

\section{Proof of Theorem~\ref{th1}}

\f\demo First, we prove $t_g(G)\leq \kappa^g(G)+g$ under the PMC and the MM model.

\begin{figure}[h!]
\begin{center}
\includegraphics[height=4cm,width=8cm]{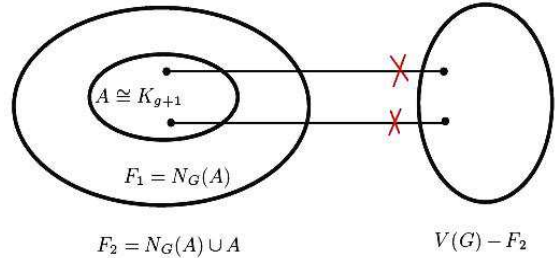}
\end{center}
\begin{center}
\vspace{-0.6cm}
\caption{ The illustration of Theorem~\ref{th1}}\label{F1}
\vspace{-0.6cm}
\end{center}
\end{figure}

Let $T$ be the minimum $g$-good-neighbor cut of $G$ satisfies Condition (1), i.e.
$G-T$ has two components, one of which is isomorphic to $K_{g+1}$, say $A$.
Clearly, $T=N_G(A)$, and $\delta(G-T)\geq g$.
Let $F_1=N_G(A)$, $F_2=N_G(A)\cup A$, see Figure~\ref{F1}. Then $|F_1|=\kappa^g(G)$,
$|F_2|=\kappa^g(G)+g+1$, $\delta(G-F_1)\geq g$ and $\delta(G-F_2)\geq g$.
It implies $F_1$ and $F_2$ are $g$-good-neighbor faulty sets of $G$. Note that $F_1\Delta F_2=A$,  $N_G(A)=F_1\subseteq F_2$, there is no edge of $G$ between $V(G)\setminus (F_1\cup F_2)$ and $F_1\Delta F_2$.
By Lemma~\ref{lem-pmc}, $G$ is not $g$-good-neighbor $(\kappa^g(G)+g+1)$-diagnosable under the PMC model,
so $t_g(G)\leq \kappa^g(G)+g$ under the PMC model.

Note that $F_1\setminus F_2=\emptyset$, $F_2\setminus F_1=A$,
$F_1$ and $F_2$ do not satisfy any one condition in Lemma~\ref{lem-mm}.
By Lemma~\ref{lem-mm}, $G$ is not $g$-good-neighbor $(\kappa^g(G)+g+1)$-diagnosable under the MM model,
so $t_g(G)\leq \kappa^g(G)+g$ under the MM model.

\medskip
Next we prove $t_g(G)\geq \kappa^g(G)+g$, i.e., $G$ is $g$-good-neighbor
$(\kappa^g(G)+g)$-diagnosable. \medskip

(I) For the PMC model, it is equivalent to prove Claim~\ref{clm1}.

\begin{clm}\label{clm1}
For each distinct pair of $g$-good-neighbor faulty sets $F_1$ and $F_2$ of $G$ with $|F_1|\leq \kappa^g(G)+g$ and $|F_2|\leq\kappa^g(G)+g$,
there is an edge $(x,y)\in E(G)$ with $x\in V(G)\setminus (F_1\cup F_2)$
and $y\in F_1\Delta F_2$.
\end{clm}

\f Proof of Claim~\ref{clm1}.
Suppose, on the contrary, that there are two distinct $g$-good-neighbor faulty sets $F_1$ and $F_2$ of $G$ with $|F_1|\leq \kappa^g(G)+g$ and $|F_2|\leq \kappa^g(G)+g$, there is no edge between $V(G)\setminus (F_1\cup F_2)$ and $F_1\Delta F_2$.

Without loss of generality, assume that $F_2\setminus F_1\neq \emptyset$.
If $V(G)=F_1 \cup F_2$, then $N=|V(G)|=|F_1\cup F_2|=|F_1|+|F_2|-|F_1\cap F_2|\leq 2\kappa^g(G)+2g<N$, it is a contradiction. Therefore, $V(G)\neq F_1 \cup F_2$.

Note that $F_1$ is a $g$-good-neighbor faulty set,
$\delta(G-F_1)\geq g$. Because there exists no edge between $V(G)\setminus(F_1\cup F_2)$ and $F_1\Delta F_2$, $\delta(G-(F_1\cup F_2))\geq g$ and $\delta(G[F_2\setminus F_1])\geq g$.
Similarly, $\delta(G[F_1\setminus F_2])\geq g$ if $F_1\setminus F_2\neq \emptyset$.
Thus, $F_1\cap F_2$ is a $g$-good-neighbor cut because of $F_2\setminus F_1\neq \emptyset$ and $G-(F_1\cup F_2)\neq \emptyset$, so $|F_1\cap F_2|\geq \kappa^g(G)$.
Note that $\delta(G[F_2\setminus F_1])\geq g$, it follows that
$|F_2\setminus F_1|\geq g+1$.
Then, $|F_2|=|F_2\setminus F_1|+|F_1\cap F_2|\geq \kappa^g(G)+g+1$,
which contradicts with $|F_2|\leq \kappa^g(G)+g$. The proof of Claim~\ref{clm1} is completed.

\medskip
(II) Now we consider the MM model. We prove $t_g(G)\geq \kappa^g(G)+g$, i.e., $G$ is $g$-good-neighbor $(\kappa^g(G)+g)$-diagnosable.

Suppose, on the contrary, that there are two distinct $g$-good-
neighbor faulty sets $F_1$ and $F_2$ of $G$ with $|F_1|\leq \kappa^g(G)+g$
and $|F_2|\leq \kappa^g(G)+g$,
but $(F_1, F_2)$ does not satisfy any one condition in Lemma~\ref{lem-mm}.
Clearly, $|F_1\cap F_2|\leq \kappa^g(G)+g-1$ because of $F_1\neq F_2$.
Without loss of generality, assume that $F_2\setminus F_1\neq \emptyset$.
If $V(G)=F_1 \cup F_2$, then $N=|V(G)|=|F_1\cup F_2|=|F_1|+|F_2|-|F_1\cap F_2|\leq 2\kappa^g(G)+2g$,
it is impossible by Condition (2).
Therefore, $V(G)\neq F_1 \cup F_2$.

\begin{clm}\label{clm2}
$G-(F_1\cup F_2)$ has no trivial component.
\end{clm}

\f Proof of Claim~\ref{clm2}.
If $g=1$, it implies that $|F_1|\leq \kappa^1(G)+1$, $|F_2|\leq\kappa^1(G)+1$
and $|F_1\cap F_2|\leq \kappa^1(G)$.
Let $W$ be the set of trivial components in $G-(F_1\cup F_2)$ and
$C=G-(F_1\cup F_2\cup W)$. Assume $|W|\neq 0$.
Then $F_1\setminus F_2\neq\emptyset$ and $F_2\setminus F_1\neq\emptyset$.
For any $w\in W$, note that $F_1$ (resp. $F_2$) is a $1$-good-neighbor faulty set,
by Lemma~\ref{lem-mm}, there is exactly one vertex $u\in F_2\setminus F_1$
(resp. $v\in F_1\setminus F_2$ ) such that $u$ (resp. $v$) is adjacent to $w$.

Note that $F_1\setminus F_2\neq \emptyset$, then
$w$ has $n-2$ neighbors in $F_1\cap F_2$, it implies that $|F_1\cap F_2|\geq n-2$.
One has $\sum\limits_{w\in W}|N_{G[F_1\cap F_2]}(w)|=|W|(n-2)\leq \sum\limits_{v\in F_1\cap F_2}d_{G}(v)=n|F_1\cap F_2|\leq n\kappa^1(G)$,
so $|W|\leq \frac{n\kappa^1(G)}{n-2}\leq 3\kappa^1(G)$.
If $C=\emptyset$, then $|V(G)|=|F_1\cup F_2|+|W|=|F_1|+|F_2|-|F_1\cap F_2|+|W|\leq 2\kappa^g(G)+3\kappa^1(G)+2g<N$
which contradicts with Condition (2).
Thus, $C\neq\emptyset$. Note that
$(F_1, F_2)$ does not satisfy the Condition (1) in Lemma~\ref{lem-mm}
and $C$ is the set of non-trivial components of $G-(F_1\cup F_2)$, so
there is no edge between $C$ and $F_1\Delta F_2$.
It implies that $F_1\cap F_2$ is a vertex-cut of $G$
and $\delta(G-(F_1\cap F_2))\geq 1$,
i.e., $F_1\cap F_2$ is a $1$-good-neighbor cut of $G$, so $|F_1\cap F_2|\geq \kappa^1(G)$.
Since $|F_1\cap F_2|\leq \kappa^1(G)$, it implies $|F_1\cap F_2|=\kappa^1(G)$.

Note that neither $F_1\setminus F_2$ nor $F_2\setminus F_1$ is empty,
so $|F_2\setminus F_1|=|F_1\setminus F_2|=1$.
Let $F_1\setminus F_2=\{v_1\}$, $F_2\setminus F_1=\{v_2\}$.
For any $w\in W$, $w$ is adjacent to both $v_1$ and $v_2$.

Note that $|F_1\cap F_2|=\kappa^1(G)$ and $F_1\cap F_2$ is a $1$-good-neighbor cut of $G$,
by Condition (3), $G-(F_1\cap F_2)$ has two components, one of which is an edge.
It follows that $v_1$ is adjacent to $v_2$ and $|W|=0$, which contradicts with $W\neq \emptyset$.

\medskip
Now we assume that $2\leq g\leq k-1$.
Since $F_1$ is a $g$-good-neighbor faulty set, for any $x\in G-F_1$,
$|N_{G-F_1}(x)|\geq g$. As the vertex set pair $(F_1, F_2)$ is not satisfied
with any one condition in Lemma~\ref{lem-mm}. By Condition (3) in Lemma~\ref{lem-mm},
any vertex $w\in V(G)\setminus(F_1\cup F_2)$ has at most one neighbor in
$F_2\setminus F_1$, it implies that $|N_{G-(F_1\cup F_2)}(w)|\geq g-1\geq 1$, i.e., $G-(F_1\cup F_2)$ has no trivial component. The Claim is completed.

\medskip
Let $y\in V(G)\setminus (F_1\cup F_2)$. By Claim~\ref{clm2},
$y$ has at least one neighbor in $G-(F_1\cup F_2)$.
Note that the vertex set pair $(F_1, F_2)$ does not satisfy any one condition in Lemma~\ref{lem-mm},
$y$ has no neighbor in $F_1\Delta F_2$.
By the arbitrary of $y$, there is no edge between $V(G)\setminus
(F_1\cup F_2)$ and $F_1\Delta F_2$.

Since $F_2\setminus F_1\neq \emptyset$, and $F_1$ is a $g$-good-neighbor faulty set and condition (3) of Lemma~\ref{lem-mm},
$\delta(G[F_2\setminus F_1])\geq g$. Similarly, $\delta(G[F_1\setminus F_2])\geq g$
if $F_1\setminus F_2\neq \emptyset$.
Since $V(G)-(F_1\cup F_2)\neq \emptyset$ and $F_2\setminus F_1\neq \emptyset$,
$F_1\cap F_2$ is a $g$-good-neighbor cut of $G$, so $|F_1\cap F_2|\geq \kappa^g(G)$.
Since $\delta(G[F_2\setminus F_1])\geq g$,
it follows that $|F_2\setminus F_1|\geq g+1$.
Then, $|F_2|=|F_2\setminus F_1|+|F_1\cap F_2|\geq \kappa^g(G)+g+1$,
which contradicts with $|F_2|\leq \kappa^g(G)+g$.
Therefore, $G$ is $g$-good-neighbor $(\kappa^g(G)+g)$-diagnosable under the MM model and $t_g(G)\geq \kappa^g(G)+g$.

By the above discussion, $t_g(G)=\kappa^g(G)+g$. The proof is completed.\hfill\qed

\section{Applications}
\subsection{Application to data center network  $D_{k,n}$}

Guo et al.~\cite{G08} proposed a server-centric data center network called DCell.
Data center networks $D_{k,n}$ have been becoming more and more important with the development of cloud computing.

Given a positive integer $m$, we use $\langle m\rangle$ and $[m]$ to denote the sets
$\{0,1,2,\ldots,m\}$ and $\{1,2,\ldots,m\}$, respectively. For any
integers $k\geq 0$ and $n\geq 2$, we use $D_{k,n}$ denote a $k$-dimensional DCell with  $n$-port switches. $D_{0,n}$ is a complete graph on $n$ vertices.
We use $t_{k,n}$ to denote the number of vertices in $D_{k,n}$ with $t_{0,n}=n$ and
$t_{i,n}=t_{i-1,n}\times (t_{i-1,n}+1)$, where $i\in [k]$.
Let $I_{0,n}=\langle n-1\rangle$ and $I_{i,n}=\langle t_{i-1,n}\rangle$
for any $i\in [k]$. Then, let $V_{k,n}=\{u_ku_{k-1}\cdots u_0:\ u_i\in I_{i,n}$
and $i\in \langle k\rangle\}$, and $V_{k,n}^{\ell}=\{u_ku_{k-1}\cdots u_\ell:\ u_i\in I_{i,n}$
and $i\in \{\ell,\ell+1,\ldots,k\}$ for any $\ell\in [k]\}$. Clearly, $|V_{k,n}|=t_{k,n}$ and
$|V_{k,n}^{\ell}|=t_{k,n}/t_{\ell-1,n}$. The definition of $D_{k,n}$ is as follows~\cite{G08}.

\begin{defi}\label{defi1}
$D_{k,n}$ is a graph with vertex set $V_{k,n}$, where a vertex $u=u_ku_{k-1}\cdots u_i\cdots u_0$ is adjacent to a vertex $v=v_kv_{k-1}\cdots v_i\cdots v_0$ if and only if there is an integer $\ell$ with
\begin{enumerate}
\item [{\rm (1)}] $u_ku_{k-1}\cdots u_\ell=v_kv_{k-1}\cdots v_\ell$,
\item [{\rm (2)}] $u_{\ell-1}\neq v_{\ell-1}$,
\item [{\rm (3)}] $u_{\ell-1}=v_0+\sum\limits_{j=1}^{\ell-2}(v_j\times t_{j-1,n})$ and $v_{\ell-1}=u_0+\sum\limits_{j=1}^{\ell-2}(u_j\times t_{j-1,n})+1$ with $\ell>1$;
\end{enumerate}

\f Or $u_k\neq v_k$, $u_k\leq v_k$ and $u_k=v_0+\sum\limits_{j=1}^{k-1}(v_j\times t_{j-1,n})$ and
$v_k=u_0+\sum\limits_{j=1}^{k-1}(u_j\times t_{j-1,n})+1$.

\end{defi}

\begin{figure}[h!]
\begin{center}
\includegraphics[height=6cm,width=7cm]{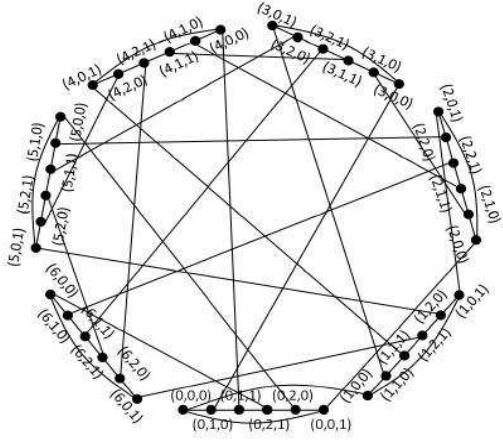}
\end{center}
\begin{center}
\vspace{-0.6cm}
\caption{ The illustration of $D_{2,2}$}\label{F2}
\vspace{-0.6cm}
\end{center}
\end{figure}

$D_{0,2}$ is an edge; $D_{1,2}$ is a cycle of length $6$. $D_{2,2}$ is shown in Figure~\ref{F2}.
It is clear that $D_{k,n}$ is a regular graph with $t_{k,n}$ vertices.

When all three conditions of Definition~\ref{defi1} hold,
we define that two adjacent vertices $u$ and $v$ have a leftmost distinct element
at position $\ell-1$. For any integer $d\geq 0$, when two adjacent vertices $u$ and $v$
have a leftmost differing element at the position $d$, denoted by ldiff$(u,v)=d$.
For any $\alpha\in V_{k,n}^{\ell}$ with $\ell\in [k]$,
we use $D_{\ell-1,n}^{\alpha}$ to denote the graph obtained by prefixing the label of each vertex of one copy of $D_{\ell-1,n}$ with $\alpha$.
Clearly, $D_{\ell-1,n}\cong D_{\ell-1,n}^{\alpha}$.
For any integers $n\geq 2$ and $k\geq1$, edges joining vertices in the same copy of $D_{k-1,n}$ are called {\it internal edges} and edges joining vertices in disjoint copies of $D_{k-1,n}$ are called {\it external edges}. Clearly, each vertex of
$D_{k-1,n}^i$ is joined to exactly one external
edge and $(n+k-2)$-internal edges for each $i\in I_{k,n}$.

\medskip

From the definition of $D_{k,n}$ in~\cite{G08}, the following properties~\ref{pro2} can be gotten directly.

\begin{prop}\label{pro2}
Let $D_{k,n}$ be the data center network  with $k\geq 0$ and $n\geq 2$.
\begin{enumerate}
\item [{\rm (1)}] $D_{0,n}$ is a complete graph with $n$ vertices labeled as $0,1,2,\ldots,n-1$ respectively.
\item [{\rm (2)}] For $k\geq 1$, $D_{k,n}$ consists of $t_{k-1,n}+1$ copies of $D_{k-1,n}$,
denoted by $D_{k-1,n}^i$, for each $i\in \langle t_{k-1,n} \rangle$.
For any two copies $D_{k-1,n}^{u_k}$ and $D_{k-1,n}^{v_k}$ of $D_{k-1,n}$ with $u_k\leq v_k$, there exists
only one edge $(u,v)$, where $u=u_ku_{k-1}u_{k-2}\cdots u_0$ in $D_{k-1,n}^{u_k}$ and $v=v_kv_{k-1}\cdots, v_0$ in $D_{k-1,n}^{v_k}$ which satisfy that $u_k=v_0+\sum\limits_{j=1}^{k-1}(v_j\times t_{j-1,n})$ and
$v_k=u_0+\sum\limits_{j=1}^{k-1}(u_j\times t_{j-1,n})+1$. It implies that each vertex in $D_{k-1,n}^{u_k}$ has only one neighbor
which is not in $D_{k-1,n}^{u_k}$, called {\it extra neighbor}.

\item [{\rm (3)}] For any two distinct vertices $u,v$ in $D_{k-1,n}^i$,
$N_{D_{k-1,n}^{I_{k,n}\setminus \{i\}}}(u)\cap N_{D_{k-1,n}^{I_{k,n}\setminus \{i\}}}(v)=\emptyset$ and $|N_{D_{k-1,n}^{I_{k,n}\setminus \{i\}}}(u)|=1$.
There is only one edge between $D_{k-1,n}^i$ and $D_{k-1,n}^j$
for any $i,j\in I_{k,n}$ and $i\neq j$.
\end{enumerate}
\end{prop}

\begin{lem}{\rm(\cite{G08})}\label{pro1}
The connectivity of $D_{k,n}$ is $\kappa(D_{k,n})=n+k-1$.
For any integers $k\geq 0$ and $n\geq 2$, the number of vertices in $D_{k,n}$
satisfies $t_{k,n}\geq (n+\frac{1}{2})^{2^k}-\frac{1}{2}$.
\end{lem}

\begin{lem}{\rm(\cite{wf16})}\label{lemd2}
For any integers $k\geq1$, $n\geq 2$, and $n-1\geq g$, if each fault-free vertex has at
least $g$ fault-free neighbor(s) in $D_{k,n}$, then there
exists a complete graph $A$ of order $g+1$ in $D_{k,n}$ such that
$N_{D_{k,n}}(A)=(g+1)(k-1)+n$, and $D_{k,n}-N_{D_{k,n}}(A)$ has exactly
two components: one is $A$ and the other is $D_{k,n}-N_{D_{k,n}}(A)-A$,
where every vertex of $D_{k,n}-N_{D_{k,n}}(A)-A$ has at least $g$
fault-free neighbor(s) in $D_{k,n}-N_{D_{k,n}}(A)-A$.
\end{lem}

\begin{lem}{\rm(\cite{wf16})}\label{lemd1}
For any integer $n\geq 2$,
$$\kappa^g(D_{k,n})=\left\{\begin{array}{ll}
(g+1)(k-1)+n& \ \text{if $0\leq g\leq n-1$ and $k\geq 1$};\\
(n+k-g-1)t_{h-n+1,n} & \ \text{if $n\leq g\leq n+k-2$ and $k\geq 2$}
\end{array}\right.$$
\end{lem}

Let $F$ be the subset of $V(D_{k,n})$. Let $F_i=F\cap D_{k-1,n}^i$, $f_i=|F_i|$ for $i\in I_{k,n}$,
$I=\{i\in I_{k,n}: f_i\geq n+k-2\}$, $F_I=\bigcup\limits_{i\in I}F_i$,
$J=I_{k,n}\setminus I$, $F_J=\bigcup\limits_{j\in J}F_j$ and $D_{k-1,n}^J=G[\bigcup\limits_{j\in J}D_{k-1,n}^j]$ which is the induced subgraph by $\bigcup\limits_{j\in J}V(D_{k-1,n}^j)$. The following Claim~\ref{clm} is useful.

\begin{clm}{\rm(\cite{wf16})}\label{clm}
Let $F$ be a faulty vertex set of $D_{k,n}$. If $|F|\leq (g+1)(k-1)+n$
with $k\geq 2$, $n\geq 2$ and $0\leq g\leq n-1$,
then $|I|\leq g+1$ and $D_{k-1,n}^J-F_J$ is connected.
\end{clm}

\begin{lem}\label{super}
$D_{k,n}$ is tightly super $(n+k-1)$-connected for $n\geq 2$ and $k\geq 2$.
\end{lem}

\f\demo Note that $\kappa(D_{k,n})=n+k-1$, let $F$ be the subset of $V(D_{k,n})$ with $|F|=n+k-1$
and $D_{k,n}-F$ is disconnected. Recall that $F_i=F\cap D_{k-1,n}^i$, $f_i=|F_i|$ for $i\in I_{k,n}$,
$I=\{i\in I_{k,n}: f_i\geq n+k-2\}$, $F_I=\bigcup\limits_{i\in I}F_i$,
$J=I_{k,n}\setminus I$, $F_J=\bigcup\limits_{j\in J}F_j$ and $D_{k-1,n}^J=G[\bigcup\limits_{j\in J}D_{k-1,n}^j]$.

By Claim~\ref{clm}, $|I|\leq 1$ and $D_{k-1,n}^J-F_J$ is connected.
We consider the following two cases.

Case 1. $|I|=0$.

In this case, $J=I_{k,n}$, $D_{k,n}-F=D_{k-1,n}^J-F_J$ is connected,
which leads to a contradiction.

Case 2. $|I|=1$.

Without loss of generality, let $I=\{1\}$, so $J=I_{k,n}\setminus \{1\}$,
$D_{k-1,n}^J-F_J$ is connected. If $D_{k-1,n}^1-F_1$ is connected, since
$|V(D_{k-1,n}^1)|=t_{k-1,n}\geq (n+\frac{1}{2})^{2^{k-1}}-\frac{1}{2}>n+2k-2=|F|$
for $n\geq 2$ and $k\geq 2$, it implies at least one vertex of $D_{k-1,n}^1-F_1$ is connected to
$D_{k-1,n}^J-F_J$. As a result, $D_{k,n}-F$ is connected, which leads to a contradiction.
In the following, assume $D_{k-1,n}^1-F_1$ is disconnected.

Subcase 2.1. $f_1=n+k-2$.

Let $u$ be the unique vertex in $F\setminus F_1$. By the similar discussion as Case 1,
$D_{k,n}-D_{k-1,n}^1-\{u\}$ is connected.
By Proposition~\ref{pro2}, any non-trivial component
of $D_{k-1,n}^1-F_1$ is connected to $D_{k,n}-D_{k-1,n}^1-\{u\}$.
There is exactly one trivial component because $|F\setminus F_1|=1$. Thus,
if $D_{k,n}-F$ is disconnected, it has
exactly two components, one of which has only one vertex, say $v$,
and its only disconnecting set is the set of the neighbors of $v$.

Subcase 2.2. $f_1=n+k-1$.

Consequently, $F_J=\emptyset$. Note that each vertex in $D_{k-1,n}^1-F_1$
is adjacent to exactly one vertex in $D_{k-1,n}^J-F_J=D_{k-1,n}^J$,
it implies $D_{k,n}-F$ is connected, which leads to a contradiction.

Hence, $D_{k,n}$ is tightly $(n+k-1)$-super connected for $n\geq 2$ and $k\geq 2$.
\hfill\qed

\begin{lem}\label{le}
Let $F\subseteq V(D_{1,n})$ and $|F|\leq n$ with $n\geq 2$. Then $D_{1,n}-F$ either is connected; or has two components, the smaller one, say $C$, $C\in \{K_t: 1\leq t\leq n\}$, where $K_t$ is the complete graph with order $t$.
\end{lem}

\f\demo If $n=2$, note that $D_{1,2}$ is a cycle of length $6$, it is not different to check the result holds. We consider $n\geq 3$ as follows.  Assume that $D_{1,n}-F$
is disconnected and $C_1,C_2,\ldots,C_m$ are the disjoint connected components of $D_{1,n}-F$.

For $i\in [m]$, $C_i$ is contained in some subgraph, say $D_{0,n}^j$ for $j\in I_{1,n}$.
If this is not true,
let $T=\{x\in I_{1,n}: C_i\cap D_{0,n}^x\neq \emptyset\}$ and $|T|\geq 2$.
Note that $D_{0,n}^x$ is a complete graph, $|V(C_i^x)|=|D_{0,n}^x|-f_x$.
As there is exactly one cross edge between $D_{0,n}^x$ and $D_{0,n}^y$,
to separate $C_i$ from other part, it has at least
$\sum\limits_{x\in T}f_x+|V(C_i)|-|T|=
\sum\limits_{x\in T}(n-|V(C_i^x)|)+|V(C_i)|-|T|=|T|(n-1)\geq 2n-2>n$
for $n\geq 3$ faulty vertices, which is a contradiction.

If $C_i\cong K_t\in D_{0,n}^j$, to separate $C_i$ from $D_{0,n}^j$, it has to remove $n-t$ vertices.
As every vertex of $C_i$ has exactly one cross edge connecting to $D_{1,n}-D_{0,n}^j$,
it need to remove $t$ vertices in $N_{D_{1,n}-D_{0,n}^j}(C_i)$, it implies there are no surplus
faulty vertices in $F$. This means that $m=1$, and $C_1\cong K_t$ is the only connected component except for the largest component in $D_{1,n}-F$.
\hfill\qed

\medskip
By Lemma~\ref{le}, $D_{1,n}$ is not tightly super $n$-connected for $n\geq 2$.

\begin{lem}\label{lemd30}
Let $F\subseteq V(D_{2,n})$ and $|F|\leq n+2$ with $n\geq 2$. Then $D_{2,n}-F$ either is connected; or has two components, one of which is a trivial component; or has two components, one of which is an edge; or has three components, two of which are trivial components.
\end{lem}

\f\demo Recall that
$I=\{i\in I_{2,n}: f_i\geq n\}$, $J=I_{2,n}\setminus I$, and $D_{1,n}^J=G[\bigcup\limits_{j\in J}D_{1,n}^j]$, by Claim~\ref{clm}, $|I|\leq2$ and $D_{1,n}^J-F_J$ is connected.
We consider the following three cases.

Case 1. $|I|=0$.

In this case, $J=I_{2,n}$, $D_{2,n}-F=D_{1,n}^J-F_J$ is connected.

Case 2. $|I|=1$.

Without loss of generality, let $I=\{1\}$, so $J=I_{2,n}\setminus \{1\}$,
$D_{1,n}^J-F_J$ is connected. If $D_{1,n}^1-F_1$ is connected, since
$|V(D_{1,n}^1)|=t_{1,n}=n(n+1)>n+2\geq|F|$
for $n\geq 2$, it implies at least one vertex of $D_{1,n}^1-F_1$ is connected to
$D_{1,n}^J-F_J$, $D_{2,n}-F$ is connected. In the following,
assume $D_{1,n}^1-F_1$ is disconnected.

Note that $f_J=|F|-f_1\leq 2$, by Proposition~\ref{pro2},
at most two vertices in $D_{1,n}^1-F_1$ are disconnected with $D_{k-1,n}^J-F_J$.
Hence, if $D_{2,n}-F$ is disconnected,
then it contains a large component and smaller components
which contain at most two vertices in total.

Case 3. $|I|=2$.

Without loss of generality, let $I=\{0,1\}$, $f_0\geq n$ and $f_1\geq n$.
Since $n+2\geq |F|\geq f_0+f_1\geq 2n$, i.e. $n\leq 2$,
so $n=2$, $f_0=2$, $f_1=2$ and $f_J=0$.

Note that $f_J=0$, any component of $D_{1,2}^i-F_i$ with more than one
vertex is adjacent to $D_{1,n}^J=D_{1,n}^J-F_J$, by Proposition~\ref{pro2}, at most one trivial component
of $D_{1,2}^i-F_i$ can be disconnected with $D_{1,n}^J-F_J$.
It leads to if $D_{2,n}-F$ is disconnected,
then it contains a large component and a trivial component.
\hfill\qed

\begin{lem}\label{lemd3}
Let $F\subseteq V(D_{k,n})$ and $|F|\leq 2k+n-2$ with $k\geq 2$ and $n\geq 2$. Then $D_{k,n}-F$ either is connected; or has two components, one of which is a trivial component; or has two components, one of which is an edge; or has three components, two of which are trivial components.
\end{lem}

\f\demo We prove the lemma by the induction on $k$.
By Lemma~\ref{lemd30}, the result holds for $k=2$.
Assume $k\geq 3$ and the result holds for $D_{k-1,n}$.
We consider $D_{k,n}$ as follows. Recall that
$I=\{i\in I_{k,n}: f_i\geq n+k-2\}$, $J=I_{k,n}\setminus I$,
and $D_{k-1,n}^J=G[\bigcup\limits_{j\in J}D_{k-1,n}^j]$,
by Claim~\ref{clm}, $|I|\leq 2$. We need only
consider the following three cases with respect to $I$.

Case 1. $|I|=0$.

In this case, $J=I_{k,n}$, $D_{k,n}-F=D_{k-1,n}^J-F_J$ is connected.

Case 2. $|I|=1$.

Without loss of generality, let $I=\{1\}$, so $J=I_{k,n}\setminus \{1\}$,
$D_{k-1,n}^J-F_J$ is connected. If $D_{k-1,n}^1-F_1$ is connected, since
$|V(D_{k-1,n}^1)|=t_{k-1,n}\geq (n+\frac{1}{2})^{2^{k-1}}>n+2k-2\geq|F|$
for $n\geq 2$ and $k\geq 3$, it implies at least one vertex of $D_{k-1,n}^1-F_1$ is connected to
$D_{k-1,n}^J-F_J$. As a result, $D_{k,n}-F$ is connected. In the following,
assume $D_{k-1,n}^1-F_1$ is disconnected.

Subcase 2.1. $n+k-2\leq f_1\leq 2k+n-4$.

By inductive hypothesis in $D_{k-1,n}^1$, if $D_{k-1,n}^1-F_1$ is disconnected,
then it contains a large component, say $B$, and smaller components
which contain at most two vertices in total.
Since $|V(D_{k-1,n}^1)|-2=t_{k-1,n}-2\geq (n+\frac{1}{2})^{2^{k-1}}-\frac{1}{2}-2>n+2k-2\geq|F|$
for $n\geq 2$ and $k\geq 3$, it implies that $B$ is connected to $D_{k-1,n}^J-F_J$.
Note that if $D_{k,n}-F$ is disconnected, then $D_{k,n}-F$ contains a
large component and smaller components
which contain at most two vertices in total.

Subcase 2.2.  $f_1=2k+n-3$.

In this case, $|F_J|=|F|-f_1\leq 1$.
Note that each vertex in $D_{k-1,n}^1$ is adjacent to exactly one vertex in $D_{k-1,n}^J$,
at most one vertex are disconnected with $D_{k-1,n}^J-F_J$.
Thus, if $D_{k,n}-F$ is disconnected, then it has two components,
one of which is a trivial component.

Subcase 2.3.  $f_1=2k+n-2$.

Consequently, $F_J=\emptyset$. Note that each vertex in $D_{k-1,n}^1-F_1$
is adjacent to exactly one vertex in $D_{k-1,n}^J-F_J=D_{k-1,n}^J$,
it leads to $D_{k,n}-F$ is connected.

Case 3. $|I|=2$.

Without loss of generality, let $I=\{1,2\}$ and $f_1\geq f_2\geq n+k-2$, so $J=I_{k,n}\setminus \{1,2\}$,
$D_{k-1,n}^J-F_J$ is connected.

We Claim $f_i=n+k-2$ for $i\in \{1,2\}$.
In fact, if $f_i\geq n+k-1$ for $i\in \{1,2\}$, then $n+2k-2\geq |F|\geq 2n+2k-2$,
it is impossible.
If $f_1=n+k-1$ and $f_2=n+k-2$, then $n+2k-2\geq |F|\geq 2n+2k-3$, i.e. $n\leq1$
it is impossible because of $n\geq 2$.

By Lemma~\ref{super}, for $i\in \{1,2\}$,
if $D_{k-1,n}^i-F_i$ is disconnected, then $D_{k-1,n}^i-F_i$ has two components,
one of which is a trivial component, say $x_i$.
Let $B_i=D_{k-1,n}^i-F_i-\{x_i\}$, $B_i$ is connected to $D_{k-1,n}^J-F_J$
by the similar discussion of Case 1. Thus, if $D_{k,n}-F$ is disconnected,
then either it has two components, one of which is a trivial component or an edge;
or has three components, two of which are trivial components.
\hfill\qed

\begin{cor}\label{cor-1}
Let $D_{k,n}$ be the data center network  with $k\geq 2$ and $n\geq 2$. Then the $g$-good neighbor
diagnosabilities of $D_{k,n}$ under the PMC model and the MM model are both $t_g(D_{k,n})=(g+1)(k-1)+n+g$ for $1\leq g\leq n-1$.
\end{cor}

\f\demo By Lemma~\ref{pro1}, $D_{k,n}$ is $(n+k-1)$-regular and $(n+k-1)$-connected and  $N=t_{k,n}\geq(n+\frac{1}{2})^{2^k}-\frac{1}{2}$. By Lemma~\ref{lemd1},
$\kappa^g(D_{k,n})=(g+1)(k-1)+n$ if $0\leq g\leq n-1$ and $k\geq 1$.
Since $N-[2\kappa^g(D_{k,n})+3\kappa^1(D_{k,n})+2g-n-1]\geq (n+\frac{1}{2})^{2^k}-\frac{1}{2}-[(g+1)(k-1)+n+3(n+2k-2)+2g-n-1]=
(n+\frac{1}{2})^{2^k}-(g+7)(k-1)-3n-2g+\frac{1}{2}>0$ for $n\geq 2$, $k\geq 2$ and $1\leq g\leq n-1$,
Condition (2) in Theorem~\ref{th1} holds;
By Lemma~\ref{lemd2}, Condition (1) in Theorem~\ref{th1} holds;
Condition (3) in Theorem~\ref{th1} holds by Lemma~\ref{lemd3}.
By Theorem~\ref{th1},  the corollary holds.
\hfill\qed

\subsection{Application to $(n,k)$-star graphs}

The $(n,k)$-star graph $S_{n,k}$, proposed by Chiang {\it et
al.}~\cite{cc95} in 1995, is another generalization of the star graph
$S_n$.

\begin{defi}\label{def-snk}
Given two positive integers $n$ and $k$ with $n>k$, let $[n]$ denote the set $\{1,2,\ldots,n\}$, and let $P_{n,k}$ be a set of
arrangements of $k$ elements in $[n]$.
The {\it $(n,k)$-star graph $S_{n,k}$} has vertex-set $P_{n,k}$, a vertex $p=p_{1}p_{2}\ldots p_{i}\ldots p_{k}$ is adjacent to a vertex
\begin{enumerate}
\item [{\rm (1)}] $p_{i}p_{2}\cdots p_{i-1}p_{1}p_{i+1}\cdots p_{k}$, where
$2\leq i\leq k$ (swap-edge).
\item [{\rm (2)}] $p'_{1} p_{2}p_{3}\cdots p_{k}$, where $p'_{1}\in
[n]\setminus \{p_{i}:\ i \in [k] \}$ (unswap-edge).
\end{enumerate}
\end{defi}

An $S_{n,k}$ can be formed by interconnecting $n$ $S_{n-1,k-1}$'s, that is,
an $S_{n,k}$ can be decomposed into $S_{n-1,k-1}$'s along any dimension
$i$, and it can also be decomposed into $n$ vertex disjoint $S_{n-1,k-1}$'s
in $k-1$ different ways by fixing one symbol in any position
$i$, $2\leq i\leq k$. We denote $S_{n,k}^i$ the subgraph which fixes the symbol $i$ in the last position $k$. Obviously, $S_{n,k}^i$ is isomorphic to $S_{n-1,k-1}$.
Moreover, there are $\frac{(n-2)!}{(n-k)!}$ independent swap-edges between $S_{n,k}^i$ and $S_{n,k}^j$ for any $i,j\in[n]$ with $i\neq j$.

Let $S_{n,k}$ be the $(n,k)$-star graph with $2\leq k\leq n-1$.
For any $\alpha=p_2p_3\cdots p_k\in P_{n,k-1}$,
let $V_{\alpha}=\{p_1\alpha:p_1\in [n]\setminus\{p_i:i\in[k]\}\}$. The the subgraph of $S_{n,k}$ induced by $V_{\alpha}$ is a complete graph of order $n-k+1$, denoted by $K_{n-k+1}^{\alpha}$.

$S_{n,k}$ is $(n-1)$-regular, $(n-1)$-connected and
vertex-transitive with order $\frac{n!}{(n-k)!}$, however, it is not edge-transitive if $n\geq k+2$ (see Chiang {\it et al.}~\cite{cc95}).
In addition, $S_{n,1}$ is isomorphic to $K_n$ and
$S_{n,n-1}$ is isomorphic to $S_n$
obviously. Moreover, Cheng {\it et al.}~\cite{cqs12} showed
$S_{n,n-2}$ is isomorphic to $AN_n$. It follows that the $(n,k)$-star graph
$S_{n,k}$ is naturally regarded as a common generalization of the
star graph $S_n$ and the alternating group network $AN_n$.


\begin{lem}{\rm(\cite{lx14})}\label{lems1}
Let $S_{n,k}$ be the $(n,k)$-star graph.
\begin{enumerate}
\item [\rm (1)] There exists a complete graph $A$ of order $g+1$ in $S_{n,k}$ such that
$N_{S_{n,k}}(A)=n+g(k-2)-1$, and $S_{n,k}-N_{S_{n,k}}(A)$ has exactly
two components: $A$ and $S_{n,k}-N_{S_{n,k}}(A)-A$, every vertex of $S_{n,k}-N_{S_{n,k}}(A)-A$ has at least $g$ fault-free neighbor(s) in $S_{n,k}-N_{S_{n,k}}(A)-A$.
\item [\rm (2)] Then $\kappa^g(S_{n,k})=n+g(k-2)-1$ for $2\leq k\leq n-1$ and $0\leq g\leq n-k$.
\end{enumerate}
\end{lem}

\begin{lem}{\rm(\cite{Z12})}\label{lem021}
Let $F$ be a faulty vertex set of $S_{n,k}$
($3\leq k\leq n-2$) with $|F|\leq n+k-3$.
Then $S_{n,k}-F$ satisfies one of the following conditions:
\begin{enumerate}
\item [\rm (1)] $S_{n,k}-F$ is connected; or
\item [\rm (2)] $S_{n,k}-F$ has two components, one of which is a trivial component; or
\item [\rm (3)] $S_{n,k}-F$ has two components, one of which ia an edge. Moreover, $F$ is formed by the neighbor of the edge.
\end{enumerate}
\end{lem}

\begin{lem}{\rm(\cite{hy95})}\label{lem022}
Let $F$ be a vertex-cut of $S_n$ for $n\geq 5$. If $|F|\leq 2n-4$, then $S_n-F$ satisfies one of the following conditions:
\begin{enumerate}
\item [{\rm (1)}] $S_n-F$ has two components, one of which is a trivial component.
\item [{\rm (2)}] $S_n-F$ has two components, one of which is an edge.
Moreover, if $|F|=2n-4$, $F$ is formed by the neighbor of the edge.
\end{enumerate}
\end{lem}

\begin{rmk}\label{r2}
Note that $S_{n,k}$ is $(n-1)$-regular $(n-1)$-connected and $|V(S_{n,k})|=\frac{n!}{(n-k)!}$.
By Lemma~\ref{lems1} (2), $\kappa^g(S_{n,k})=n+g(k-2)-1$ for $3\leq k\leq n-1$ and $0\leq g\leq n-k$.
Since $N-[2\kappa^g(S_{n,k})+3\kappa^1(S_{n,k})+2g-n-1]>0$ for $3\leq k\leq n-1$ and $1\leq g\leq n-k$, Condition (2) in Theorem~\ref{th1} holds;
By Lemma~\ref{lem021} and Lemma~\ref{lem022}, Condition (3) in Theorem~\ref{th1} holds;
By Lemma~\ref{lems1} (1), Condition (1) in Theorem~\ref{th1} holds;
By Theorem~\ref{th1}, we can deduce the following Corollary holds.
\end{rmk}

\begin{cor}{\rm(\cite{xz16})}\label{cor-2}
Let $S_{n,k}$ be the $(n,k)$-star graph with $3\leq k\leq n-1$.
Then the $g$-good neighbor diagnosabilities of $S_{n,k}$ under the PMC model and the MM model are both $t_g(S_{n,k})=n+g(k-1)-1$ for $1\leq g\leq n-k$.
\end{cor}

\medskip
Since the star graph $S_n$ is isomorphic to $S_{n,n-1}$ and the alternating group network $AN_n$ is isomorphic to $S_{n,n-1}$~\cite{cqs12}.
The following corollaries are obtained directly from Corollary~\ref{cor-2}.

\begin{cor}\label{cor1}
Let $S_n$ be the $n$-dimensional star graphs for $n\geq 4$.
Then $1$-good-neighbor diagnosabilities of $S_n$ under the two models are both $2n-3$.
\end{cor}

\begin{cor}\label{cor2}
Let $AN_n$ be the $n$-dimensional alternating group network for $n\geq 4$.
Then $g$-good-neighbor diagnosabilities of $AN_n$ under the two models
are both $t_g(AN_n)=n+g(n-2)-1$ for $1\leq g\leq 2$ and $n\geq 4$.
\end{cor}

\subsection{Application to $(n,k)$-arrangement graphs}

The {\it $(n,k)$-arrangement graph}, denoted by $A_{n,k}$, was
proposed by Day and Tripathi~\cite{dt92} in 1992. The definition of
$A_{n,k}$ is as follows.

\begin{defi}\label{defi-1}
Given two positive integers $n$ and $k$ with $n>k$, let $[n]$ denote the set $\{1,2,\ldots,n\}$, and let $P_{n,k}$ be a set of
arrangements of $k$ elements in $[n]$. The $(n,k)$-arrangement graph, denoted by $A_{n,k}$, has vertex-set $P_{n,k}$ and two
vertices are adjacent if and only if they differ in exactly one
position.
\end{defi}

$A_{n,k}$ is $k(n-k)$-regular, $k(n-k)$-connected with $\frac{n!}{(n-k)!}$ vertices,
vertex-transitive and edge-transitive (see~\cite{dt92}). Clearly,
$A_{n,1}$ is isomorphic to the complete graph $K_n$ and $A_{n,n-1}$ is isomorphic to the $n$-dimensional star graph $S_n$. Chiang and
Chen~\cite{cc98} showed that $A_{n,n-2}$ is isomorphic to
the $n$-alternating group graph $AG_n$.

For a fixed $i$ ($1\leqslant i\leqslant k)$, let
 $$
 V_i=\{p_1\cdots p_{i-1}q_ip_{i+1}\cdots p_k:\ q_i\in I_n\setminus\{p_1,
 \cdots, p_{i-1}, p_{i+1}, \cdots, p_k\}\}
$$
Then $|V_i|=n-k+1$. There are $|P_{n,k-1}|$ such $V_i$'s. By
definition, it is easy to see that the subgraph of $A_{n,k}$ induced
by $V_i$ is a complete graph $K_{n-k+1}$. In special,
$K_{n-k+1}=K_n$ if $k=1$, and $K_{n-k+1}=K_2$ if $k=n-1$.
Thus, when $n\geqslant k+2$ and $k\geqslant 2$, for each fixed $i$
($1\leqslant i\leqslant k)$, the vertex-set of $A_{n,k}$ can be
partitioned into $|P_{n,k-1}|$ subsets, each of which induces a
complete graph $K_{n-k+1}$.


\begin{lem}{\rm(\cite{lz12})}\label{lema1}
Let $A_{n,k}$ be the $(n,k)$-arrangement graph.
\begin{enumerate}
\item [\rm (1)] There exists a complete graph $A$ of order $g+1$ in $A_{n,k}$ such that
$N_{A_{n,k}}(A)=[(g+1)k-g](n-k)-g$, and $A_{n,k}-N_{A_{n,k}}(A)$ has exactly
two components: one is $A$ and the other is $A_{n,k}-N_{A_{n,k}}(A)-A$,
where every vertex of $A_{n,k}-N_{A_{n,k}}(A)-A$ has at least $g$
fault-free neighbor(s) in $A_{n,k}-N_{A_{n,k}}(A)-A$.
\item [\rm (2)] $\kappa^g(A_{n,k})=[(g+1)k-g](n-k)-g$
for $3\leq k\leq n-1$ and $1\leq g\leq \min\{k-2,n-k\}$.
\end{enumerate}
\end{lem}

\begin{lem}{\rm(\cite{Z1})}\label{lema2}
Let $F$ be a set of faulty vertices in $A_{n,k}$ with $|F|\leq (2k-1)(n-k)-1$,
and $k\geq 3$. If $A_{n,k}-F$ is disconnected, then it
has exactly two components, one of which is a single vertex or a single edge.
Moreover, if $|F|=(2k-1)(n-k)-1$, $F$ is formed by the neighbors of the edge.
\end{lem}

\begin{rmk}\label{r3}

Note that $A_{n,k}$ is $k(n-k)$-regular $k(n-k)$-connected and $N=|V(A_{n,k})|=\frac{n!}{(n-k)!}$.
By Lemma~\ref{lema1}, $\kappa^g(A_{n,k})=[(g+1)k-g](n-k)-g$
for $3\leq k\leq n-1$ and $1\leq g\leq \min\{k-2,n-k\}$, Condition (1) holds.
Since $N-[2\kappa^g(A_{n,k})+3\kappa^1(A_{n,k})+2g-n-1]>0$ for $3\leq k\leq n-1$ and $1\leq g\leq \min\{k-2,n-k\}$, Condition (2) in Theorem~\ref{th1} holds;
Condition (3) in Theorem~\ref{th1} holds by Lemma~\ref{lema2}.
Thus, by Theorem~\ref{th1}, we can deduce the following Corollary holds.
\end{rmk}

\begin{cor}\label{cor-3}
Let $A_{n,k}$ be the $(n,k)$-arrangement graph. Then the $g$-good neighbor
diagnosabilities of $A_{n,k}$ under the PMC model and the MM model are both $t_g(A_{n,k})=[(g+1)k-g](n-k)$ for $3\leq k\leq n-1$ and $1\leq g\leq \min\{k-2,n-k\}$.
\end{cor}

\medskip
Since the alternating group graph $AG_n$ is isomorphic to $A_{n,n-2}$~\cite{cc98}.
The following corollary is derived directly from Corollary~\ref{cor-3}.

\begin{cor}\label{cor3}
Let $AG_n$ be the $n$-dimensional alternating group network for $n\geq 4$.
Then $g$-good-neighbor diagnosabilities of $AG_n$ under the two models
are both $t_g(AG_n)=2[(g+1)(n-2)-g]$ for $1\leq g\leq 2$.
\end{cor}

\section{Conclusion}
$R^g$-connectivity $\kappa^g(G)$ and $g$-good-neighbor diagnosability $t_g(G)$
are two metrics to evaluate a multiprocessor system.
In this paper, we firstly established the relation between $g$-good-neighbor diagnosability and
$R^g$-connectivity for regular graphs. Secondly, we prove that $D_{k,n}$ is tightly super $(n+k-1)$-connected for $n\geq 2$ and $k\geq 2$, but $D_{1,n}$ is not tightly super $n$-connected.
Thirdly, we show that the $g$-good-neighbor diagnosability of $D_{k,n}$ are $t_g(D_{k,n})=(g+1)(k-1)+n+g$ for $1\leq g\leq n-1$ under the PMC model and the MM model, respectively.
As direct corollaries, the $g$-good-neighbor diagnosability
of the $(n,k)$-star networks $S_{n,k}$ and the $(n,k)$-arrangement graphs $A_{n,k}$ are obtained.
This method can be used to other complex networks.

\section*{Acknowledgments}

This work was supported by the National Natural Science Foundation of China
(No. 11371052, No.11571035, No.11271012, No.61572010), the Fundamental Research Funds for the Central Universities (Nos. 2016JBM071, 2016JBZ012) and the $111$ Project of China (B16002).


\end{document}